\newcommand{\version}{March 30, 2001}

\documentclass[12pt]{article}
\usepackage{a4,amsthm,amsfonts,latexsym,amssymb}
\usepackage{curves}

\linethickness{0.15mm}
\renewcommand\diskpitchstretch{2}

\swapnumbers
\theoremstyle{plain}
\newtheorem{thm}{THEOREM}
\newtheorem{cl}[thm]{COROLLARY}
\newtheorem{lem}[thm]{LEMMA}
\newtheorem{define}[thm]{DEFINITION}
\newtheorem{proposition}[thm]{PROPOSITION}
\newtheorem{conjecture}[thm]{CONJECTURE}
\theoremstyle{definition}

\newcommand{\beq}{\begin{equation}}
\newcommand{\eeq}{\end{equation}}
\def\beqa{\begin{eqnarray}}
\def\eeqa{\end{eqnarray}}
\newcommand{\half}{\mbox{$\frac{1}{2}$}}

\newcommand{\Tr}{{\rm Tr}} 

\newcommand{\A}{{\mathcal A}}
\newcommand{\Hh}{{\mathcal H}}
\newcommand{\M}{{\mathcal M}}
\newcommand{\F}{{\mathcal F}}
\newcommand{\G}{{\mathcal G}}
\newcommand{\oo}{{\mathcal O}}
\newcommand{\bb}{{\mathbf b}}
\newcommand{\Z}{{\mathbb Z}}
\newcommand{\C}{{\mathbb C}}
\newcommand{\one}{{\mathbb I}}

\newcommand{\R}{{\mathbb R}}

\newcommand{\x}{{\bf x}}
\newcommand{\aaa}{{\bf a}}
\newcommand{\cc}{{\bf c}}

\date{\small\version}

\begin{document}
\markboth{\scriptsize{BB \version}}{\scriptsize{BB \version}}

\title{\bf{A partial ordering of sets, making mean entropy monotone}}
\author{\vspace{5pt} Bernhard Baumgartner$^1$\\
\vspace{-4pt}\small{Institut f\"ur Theoretische Physik, Universit\"at Wien}\\
\small{Boltzmanngasse 5, A-1090 Vienna, Austria}}

\maketitle

\begin{abstract}

Consider a state of a system with several subsystems. The entropies of the reduced state 
on different subsystems obey certain inequalities, provided there is an 
equivalence relation, and a function measuring volumes or weights of subsystems. 
The entropy per unit volume or unit weight, the mean entropy, is then decreasing with 
respect to an order relation of the subsystems, defined in this paper. In the 
context of statistical mechanics a lattice system is studied in detail, 
and a decrease of mean energy is deduced for blow-up sequences of regular and 
irregular octogons. 
\\[3ex] 
PACS numbers: \qquad 05.50.+q, \quad 03.67.-a, \quad 02.10.Ab

\end{abstract}

\footnotetext[1]{E-Mail:
\texttt{baumgart@ap.univie.ac.at}}

\section{Introduction}
{\it Entropy} is a key concept in several areas, from thermodynamics to dynamical systems. 
The study presented here refers primarily to statistical mechanics, but it may also be 
considered in the context of information and quantum information theory. 
See, for example, \cite{NC00} for the basic setup needed to implement results of this paper. 
The problem studied in this paper, {\it comparing entropies of subsystems}, 
is presented in the context of statistical mechanics first, 
then it is stated in a formulation refering to (quantum) information theory. 
Both presentations are rather abstract, but yet not as general as would be possible. 
It turns out that the methods which will be used are very flexible, 
so they may be applied also under different cicumstances. 
This will be remarked at some places. 

Consider a quantum {\it lattice system}, defined with a state on the inductive limit 
of local algebras of observables. Assume that a density matrix and an entropy $S(A)$ is assigned to 
each restriction of this state to any local algebra $\A _A$ belonging to a 
finite subset $A$ of the lattice. If the state is invariant under translations, 
and if $B$ is a {\it ``translate''} of $A$ 
(meaning, that there exists a translation, mapping $A$ to $B$), we know that $S(B)=S(A)$. 
Now we are interested in relations between entropies assigned to different sets, 
with possibly different measures $\mu (A)$. 
One knows for some time already, see Propositions 6.2.25 and 
6.2.38 in \cite{BR81}, and \cite{ABK01} with the references quoted there, 
of a subadditivity, a strong subadditivity, a triangle inequality, 
and a ``strong triangle inequality'' for entropies,   
and of the existence of the van Hove limit of the {\it mean entropy}, the entropy per unit volume 
(also called ``entropy density'') 
\beq \label{mean}
s(A)=S(A)/\mu (A).
\eeq
All these assertions are true 
in case the local algebras $\A_A$ are represented on local Hilbert spaces $\Hh_ A$ 
with the product property
\beq \label{product}
\Hh _D = \Hh _A \otimes \Hh_C, \qquad if \quad D=A\cup C \quad and \quad A\cap C=\emptyset , 
\eeq
and the density matrices fulfill the compatibility condition 
\beq \label{compati}
\rho_A=\Tr_C\rho_D, \qquad if \quad D=A\cup C \quad and \quad A\cap C=\emptyset ,
\eeq
where $\rho_A$ represents the state on the local algebra $\A_A$. 
See \cite{LR73a} for a listing of fundamental properties of entropy related 
to the product property of algebras, from which all the above assertions follow. 
See also \cite{W78} for properties of entropy, and for a proof of encrease of entropy 
of lines of increasing length in one dimensional translation invariant systems. 

Then it has been established for mean entropy in \cite{ABK98} that $s(A) \geq s(B)$, 
if $A$ and $B$ are rectangular boxes or parallelepipeds with $A \subset B$. 
In the same electronic ``paper'' the question has been raised, whether this relation of monotonicity 
can be extended to {\it all} pairs of sets with $A \subset B$. 
Here in the present paper, in the Appendix, we will answer this question in the negative, 
by giving counterexamples, considering systems with spins on $\Z^2$. 
But we can nevertheless make some progress concerning ``Question 1'' of \cite{ABK01}: 
We extend the monotonicity property of $s(A)$ from rectangular boxes to more general volumes. 

To enable this quest we define an {\it order relation between sets}, adapted to imply monotone 
decrease of mean entropy. 
The formalism of this method 
is not tied to lattice systems. All we need is an equivalence relation between sets, 
corresponding to an invariance property of the state, as for example translation 
invariance. To define the mean entropy we need of course a measure. 
Then we will need nothing more but the strong subadditivity of the entropy 
as a function of sets,
\beq \label{ssubad}
S(A \cup B) \leq S(A) +S(B) - S(A \cap B),
\eeq
which is a consequence of the strong subadditivity proven in \cite{LR73b}, 
and of the product property (\ref{product}).
(It is also true for fermionic systems with a modification of the product property, \cite{AM01}.)
We remark that we do not need positivity of the entropy. So our results are true for 
classical continuous systems as well.

Refering to quantum {\it information}, consider information stored in an assembly of several 
elementary quantum systems. Consider also subsystems of the total system, 
defined by subsets of the total set of elementary systems. Each subsystem $A$, and the 
information it carries, is represented on a Hilbert space $\Hh _A$ with a density 
matrix $\rho _A$, obeying the product property (\ref{product}) and the 
compatibility condition (\ref{compati}). To any subsystem $A$ 
a weight $\mu (A)$ is assigned. These weights fulfill the conditions for a measure on 
the family of subsets $\{A,B,\ldots \}$: $0\leq \mu (A) \leq 1$, 
$\mu (A\cup B)+\mu (A\cap B)=\mu (A)+ \mu (B)$. 
Assume moreover, that this system is equipped with an equivalence relation on the 
family of subsets, which is compatible with the assignement of a 
mean entropy, the entropy per weight, so that 
``$A'$ equivalent to $A$'' implies $S(A')=S(A)$. (It is not necessarely compatible with 
the assignement of a weight.) Now we are interested not only in the von Neumann entropy 
$S(A)$, but also in the {\it mean entropy}, the entropy per unit weight, defined in (\ref{mean}). 
When can we be sure, that $s(A)\geq s(B)$? Also this question can be answered 
with the help of the order relation to be defined in section \ref{order}, 
using nothing more but the strong subadditivity (\ref{ssubad}). 

The definition of partial ordering is given in axiomatic form, with axioms 
defining the process how to construct the order relation step by step.  
We take some care of the logic in this process: 
Examples demonstrate both the use of these axioms, using subsets of a two-dimensional lattice, 
and their mutual logical independence, that is, that no axiom is a consequence of the others. 
Each step in the creation of such an order relation is consistent with {\it monotonicity} 
of the mean entropy. This fact is a consequence of strong superadditivity (\ref{ssubad}); 
it is stated and proven in Theorem \ref{monotoneth} and its proof, in section \ref{monmean}. 
Finally, we present an application of this method in the context 
of statistical mechanics, creating an
order relation for convex octogons in $\Z^2$. Their ordering is related to simple 
inequalities for the pieces of the boundaries, as stated in Definition \ref{orderbd}
and in Theorem \ref{octogonorder}. 
It includes the ordering of {\it blow up} sequences of octogons, and implies the 
monotone decrease of mean entropy.

\section{The partial ordering}\label{order}
Consider a {\it space} (set) $\M$, with a family $\F$ of subsets which is 
closed under forming unions and intersections of pairs, and an equivalence 
relation $\sim$ on $\F$. 
A partial ordering $\prec$ on $\F$ is defined by the following {\bf axioms}:

\begin{enumerate}
\item[(I)] If $A\prec B$, $A\sim A'$ and $B\sim B'$, then also 
$A'\prec B$ and $A\prec B'$ hold. ({\bf Invariance})
\item[(II)] If $A\cap B$ is empty, and if either 
$A\sim B$ or $A\prec B$ holds, then $A\prec D=A\cup B$.
\item[(III)] If $A\cap B =C\prec B$, and if either 
$A\sim B$ or $A\prec B$ holds, then $A\prec D=A\cup B$.
\item[(IV)] If $A\cap B =C \prec D=A\cup B$, and if either 
$A\sim B$ or $A\prec B$ holds, then $A\prec D$.
\item[(V)] If $B\cap C=A\prec B$ and also $A\prec C$ hold, then $A\prec D=B\cup C$, 
even if $B$ is not comparable with $C$. 
(i.e: Neither $B\sim C$, nor $B\prec C$, nor $C\prec B$ hold.) 
\item [(VI)] If $A\prec B$ and $B\prec D$, then also $A\prec D$ holds. ({\bf Transitivity})
\end{enumerate}

To enable a precise referencing, we subdivide the second, the third and the fourth axiom:

\begin{enumerate}
\item[(II)]
\begin{enumerate}
\item[(a)] $A\cap B =\emptyset$ and $A\sim B$ \quad $\Rightarrow$ \quad $A\prec D=A\cup B$
\item[(b)] $A\cap B =\emptyset$ and $A\prec B$ \quad $\Rightarrow$ \quad $A\prec D=A\cup B$
\end{enumerate}
\item[(III)]
\begin{enumerate}
\item[(a)] $A\cap B =C\prec B$ and $A\sim B$ \quad $\Rightarrow$ \quad $A\prec D=A\cup B$
\item[(b)] $A\cap B =C\prec B$ and $A\prec B$ \quad $\Rightarrow$ \quad $A\prec D=A\cup B$
\end{enumerate}
\item[(IV)]
\begin{enumerate}
\item[(a)] $A\cap B =C\prec D=A\cup B$ and $A\sim B$ \quad $\Rightarrow$ \quad $A\prec D$
\item[(b)] $A\cap B =C\prec D=A\cup B$ and $A\prec B$ \quad $\Rightarrow$ \quad $A\prec D$
\end{enumerate}
\end{enumerate}

The condition of uncomparability is included in the fifth axiom to keep the axioms 
mutually independent; but this condition can be ignored in applications: 
\begin{lem}\label{axiom5}
Let $A=B\cap C$, $A\prec B$, $A\prec C$ and either $B\sim C$ or 
$B\prec C$. Then  $A\prec D=B\cup D$.
\end{lem}
\begin{proof}
Use axioms IIIa and VI if $B\sim C$, axioms IIIb and VI if $B\prec C$.
\end{proof}
On the other hand, allowing equivalence of $B$ with $C$ in axiom V would make the 
order relation stated in axiom IIIa a consequence 
of axioms I, V and VI.

There are in principle two different ways in using these axioms. One possibility 
would be to ``import'' an order relation between sets, ``dictating'' it, and then extending it. 
In this paper however we are concerned with the other way of starting from scratch, 
{\bf creating} the ordering out of the equivalence relation, by forming unions and 
checking for order relations concerning the intersections.
We state an immediate consequence of using this creative process:
\begin{lem}\label{unionlem}
If the ordering is created in a countable number of steps, then $A\prec D$ implies that 
there exist sets $A^{(1)}...A^{(N)}$ which are equivalent to $A$, such that 
\beq\label{union}
D=\bigcup^N_{n=1} A^{(n)}
\eeq
\end{lem}
\begin{proof}
We may order the sequence of logical steps leading to $A\prec D$ 
in such a way, that in the n'th step, establishing $...\prec B^{(n)}$, only the relations 
established in the former steps are needed. 
Going backwards from $A\prec D$, we see that $D$ and each $B^{(n)}$ has to be 
represented as the union of two sets, $B^{(\nu)}$ and $B^{(\mu)}$, which are either equivalent to $A$ 
or are ordered like $A\prec B^{(\nu)}$.
Then we proceed by induction. 
The basis of the creative process is the axiom IIa. It is the only one, which does 
not need a pair of sets which are already in an order relation. So $B^{(1)}=A^{(1)}\cup A^{(2)}$.
We assume that $A\prec B^{(\nu)}$ for $\nu < n-1$ has been established.  
Now either a similar procedure as in step one may occur in step n, or  
the existence of one or two of the former relations  
is needed. Checking the axioms it is obvious that the property (\ref{union}) propagates 
from the $B^{(\nu)}$ to $B^{(n)}$ in the process of establishing $A\prec B^{(n)}$.
\end{proof}
It is to be remarked, that (\ref{union}) is not a sufficient condition for $A\prec D$. 
In the Appendix we will present a counterexample, combined with $s(A)\not\geq s(D)$. 

The following may in applications apply to the empty set: 
\begin{cl}
If a set $Z$ is not equivalent to any other set, then no relation $Z\prec D$ can be established.
\end{cl}

\section{Logical independence of the axioms and examples of applications}\label{examples}

The axiom of invariance has a special status, just enabling full power for the other 
axioms in creating the order relation. Especially axioms 
IIb and IVb would have no meaning at all without this invariance. 
Without it, order relations can only be established between sets where one 
of them is a subset of the other one. This can be seen by inspection of the ``creative'' 
axioms, II to V. So the independence of this axiom is obvious. 

Now we demonstrate the independence of the other axioms, II to VI, 
- with one exception - on examples for each one of them, 
where an ordered pair would {\it not} be comparable without it, i.e. we present each time 
an order relation $A\prec D$, which can not be created without this axiom. 
The exception is axiom IIb. If the ordering is created in a countable number of steps, 
starting with axiom IIa, then the assertion of axiom IIb is a consequence of axiom V 
in its generalized form. 
This strange fact is now being stated as Theorem \ref{twobe}. 
Nevertheless we will present two simple examples of how to use axiom IIb.
\begin{lem}[Deduction of IIb]\label{twobe}
If the ordering is created as stated above by the axioms I...VI, 
the assertion of axiom IIb is a consequence 
of the other axioms.
\end{lem}
\begin{proof}
Axiom IIb assumes $A\prec B$ and $A\cap B=\emptyset$. By Lemma \ref{unionlem}, $B$ is a union 
of sets $A^{(n)}$ which are all eqivalent to $A$. Now define $C=A\cup A^{(1)}$, observe 
$A\prec C$ by IIa, and use Lemma \ref{axiom5}.
\end{proof}

\subsubsection*{Examples}

In our system 
the two-dimensional lattice $\Z^2$ is the space $\M$, the finite subsets of $\Z^2$ 
form the family $\F$, and the discrete group of two-dimensional translations maps 
each set $A$ to all equivalent sets $A'\sim A$. 
To make the notation short, we denote those points of $\Z^2$ we need 
like the squares of a chessboard, i.e. $a1$ instead of $(0,0)$, $c4$ instead of $(2,3)$ etc. 
In the pictorial presentation of these examples the points are represented 
by squares, and the point $a1$ (which may be also outside of $A$) is marked with a cross.

\begin{enumerate}
\item[(IIa)]
$A=\begin{picture}(10,10)(0,0)
  \curve(3,3, 7,7)
  \curve(3,7, 7,3)
  \curve(0,0, 10,0)
  \curve(0,0, 0,10)
  \curve(10,0, 10,10)
  \curve(0,10, 10,10)
\end {picture}
\prec
D=\begin{picture}(20,10)(0,0)
  \curve(3,3, 7,7)
  \curve(3,7, 7,3)
  \curve(0,0, 20,0)
  \curve(0,10, 20,10)
  \curve(0,0, 0,10)
  \curve(10,0, 10,10)
  \curve(20,0, 20,10)
\end {picture}$\hspace{1pt}; \qquad
also \quad
 $\begin{picture}(10,10)(0,0)
  \curve(3,3, 7,7)
  \curve(3,7, 7,3)
  \curve(0,0, 10,0)
  \curve(0,0, 0,10)
  \curve(10,0, 10,10)
  \curve(0,10, 10,10)
\end {picture}
\prec
  \begin{picture}(10,20)(0,0)
  \curve(3,3, 7,7)
  \curve(3,7, 7,3)
  \curve(0,0, 10,0)
  \curve(0,10, 10,10)
  \curve(0,20, 10,20)
  \curve(0,0, 0,20)
  \curve(10,0, 10,20)
\end {picture}$\hspace{1pt}; \qquad
and also \quad
 $\begin{picture}(20,10)(0,0)
  \curve(3,3, 7,7)
  \curve(3,7, 7,3)
  \curve(0,0, 20,0)
  \curve(0,10, 20,10)
  \curve(0,0, 0,10)
  \curve(10,0, 10,10)
  \curve(20,0, 20,10)
\end {picture}
\prec
 \begin{picture}(20,20)(0,0)
  \curve(3,3, 7,7)
  \curve(3,7, 7,3)
  \curve(0,0, 20,0)
  \curve(0,10, 20,10)
  \curve(0,20, 20,20)
  \curve(0,0, 0,20)
  \curve(10,0, 10,20)
  \curve(20,0, 20,20)
 \end {picture}$\newline
 $A=\{a1\}\sim B=\{b1\}$, so \newline
   $A\prec D=A\cup B=\{a1,b1\}$; \newline also $\{a1\} \prec \{a1,a2\}$; \qquad and also 
   $\{a1,b1\} \prec \{a1,b1,a2,b2\}$  
 \item[(IIb)] 
$A=\quad\begin{picture}(10,20)(0,0)
  \curve(3,3, 7,7)
  \curve(3,7, 7,3)
  \curve(0,10, 10,10)
  \curve(0,20, 10,20)
  \curve(0,10, 0,20)
  \curve(10,10, 10,20)
 \end {picture}\quad
\prec
D=\quad\begin{picture}(20,20)(0,0)
  \curve(3,3, 7,7)
  \curve(3,7, 7,3)
  \curve(0,0, 20,0)
  \curve(0,10, 20,10)
  \curve(0,20, 10,20)
  \curve(0,0, 0,20)
  \curve(10,0, 10,20)
  \curve(20,0, 20,10)
\end {picture}$\hspace{1pt}; \qquad
also 
 $\quad\begin{picture}(20,20)(0,0)
  \curve(3,3, 7,7)
  \curve(3,7, 7,3)
  \curve(0,10, 20,10)
  \curve(0,20, 20,20)
  \curve(0,10, 0,20)
  \curve(10,10, 10,20)
  \curve(20,10, 20,20)
 \end {picture}\quad
\prec \quad
 \begin{picture}(30,20)(0,0)
  \curve(3,3, 7,7)
  \curve(3,7, 7,3)
  \curve(0,0, 30,0)
  \curve(0,10, 30,10)
  \curve(0,20, 20,20)
  \curve(0,0, 0,20)
  \curve(10,0, 10,20)
  \curve(20,0, 20,20)
  \curve(30,0, 30,10)
 \end {picture}$ \newline
$A=\{a2\} \prec D$ \newline $D=\{a1,b1,a2\}$, \newline 
with $B=\{a1,b1\}$, \qquad \newline and $A\prec B$ by the order created above
  and the invariance property. \newline 
  Also $A=\{a2,b2\} \prec D=\{a1,b1,c1,a2,b2\}$, with $B=\{a1,b1,c1\}$ and the 
  ordering $A\prec B$ created in the following example. 

\item[(IIIa)] 
$A=\begin{picture}(20,10)(0,0)
  \curve(3,3, 7,7)
  \curve(3,7, 7,3)
  \curve(0,0, 20,0)
  \curve(0,10, 20,10)
  \curve(0,0, 0,10)
  \curve(10,0, 10,10)
  \curve(20,0, 20,10)
 \end {picture}
\prec
 D=\begin{picture}(30,10)(0,0)
  \curve(3,3, 7,7)
  \curve(3,7, 7,3)
  \curve(0,0, 30,0)
  \curve(0,10, 30,10)
  \curve(0,0, 0,10)
  \curve(10,0, 10,10)
  \curve(20,0, 20,10)
  \curve(30,0, 30,10)
 \end {picture}\qquad$ with
 $\quad B=\begin{picture}(30,10)(0,0)
  \curve(3,3, 7,7)
  \curve(3,7, 7,3)
  \curve(10,0, 30,0)
  \curve(10,10, 30,10)
  \curve(10,0, 10,10)
  \curve(20,0, 20,10)
  \curve(30,0, 30,10)
 \end {picture}\qquad$ and
 $\quad C=\begin{picture}(20,10)(0,0)
  \curve(3,3, 7,7)
  \curve(3,7, 7,3)
  \curve(10,0, 20,0)
  \curve(10,10, 20,10)
  \curve(10,0, 10,10)
  \curve(20,0, 20,10)
 \end {picture}$ \newline
 $A=\{a1,b1\} \prec D$ \newline $D=\{a1,b1,c1\}$, with \newline $B=\{b1,c1\}$, \qquad $C=\{b1\}$,
 \newline and with the order $C\prec A\sim B$ created with IIa. 
\item[(IIIb)] 
 $A=\quad \begin{picture}(20,30)(0,0)
  \curve(3,3, 7,7)
  \curve(3,7, 7,3)
  \curve(0,10, 20,10)
  \curve(0,20, 20,20)
  \curve(0,30, 10,30)
  \curve(0,10, 0,30)
  \curve(10,10, 10,30)
  \curve(20,10, 20,20)
 \end {picture}\quad
\prec
 D=\begin{picture}(30,30)(0,0)
  \curve(3,3, 7,7)
  \curve(3,7, 7,3)
  \curve(0,0, 30,0)
  \curve(0,10, 30,10)
  \curve(0,20, 20,20)
  \curve(0,30, 10,30)
  \curve(0,0, 0,30)
  \curve(10,0, 10,30)
  \curve(20,0, 20,20)
  \curve(30,0, 30,10)
 \end {picture}$\qquad
with 
 $\quad B=\begin{picture}(30,20)(0,0)
  \curve(3,3, 7,7)
  \curve(3,7, 7,3)
  \curve(0,0, 30,0)
  \curve(0,10, 30,10)
  \curve(0,20, 20,20)
  \curve(0,0, 0,20)
  \curve(10,0, 10,20)
  \curve(20,0, 20,20)
  \curve(30,0, 30,10)
 \end {picture}\qquad $ and \quad
 $C=\begin{picture}(20,20)(0,0)
  \curve(3,3, 7,7)
  \curve(3,7, 7,3)
  \curve(0,10, 20,10)
  \curve(0,20, 20,20)
  \curve(0,10, 0,20)
  \curve(10,10, 10,20)
  \curve(20,10, 20,20)
 \end {picture}$
  \newline $A=\{a1,b1,a2\} \prec D$ \newline 
  $D=\{a1,b1,c1,a2,b2,a3\}$, with \newline $B=\{a1,b1,c1,a2,b2\}$, \newline
  $C=\{a2,b2\}$, \newline and the order $C\prec B$ created with IIb and IIIa. 
 \item[(IVa)] 
 $A=\begin{picture}(30,20)(0,0)
  \curve(3,3, 7,7)
  \curve(3,7, 7,3)
  \curve(0,0, 30,0)
  \curve(0,10, 30,10)
  \curve(10,20, 20,20)
  \curve(0,0, 0,10)
  \curve(10,0, 10,20)
  \curve(20,0, 20,20)
  \curve(30,0, 30,10)
 \end {picture}
 \prec
 D=\begin{picture}(40,20)(0,0)
  \curve(3,3, 7,7)
  \curve(3,7, 7,3)
  \curve(0,0, 40,0)
  \curve(0,10, 40,10)
  \curve(10,20, 30,20)
  \curve(0,0, 0,10)
  \curve(10,0, 10,20)
  \curve(20,0, 20,20)
  \curve(30,0, 30,20)
  \curve(40,0, 40,10)
 \end {picture}$ \qquad
with
 $\quad B=\begin{picture}(40,20)(0,0)
  \curve(3,3, 7,7)
  \curve(3,7, 7,3)
  \curve(10,0, 40,0)
  \curve(10,10, 40,10)
  \curve(20,20, 30,20)
  \curve(10,0, 10,10)
  \curve(20,0, 20,20)
  \curve(30,0, 30,20)
  \curve(40,0, 40,10)
 \end {picture}\qquad$
and
 $\quad C=\begin{picture}(20,20)(0,0)
  \curve(3,3, 7,7)
  \curve(3,7, 7,3)
  \curve(10,0, 30,0)
  \curve(10,10, 30,10)
  \curve(10,0, 10,10)
  \curve(20,0, 20,10)
  \curve(30,0, 30,10)
 \end {picture}$ \newline
$A=\{a1,b1,c1,b2\} \prec D$ \newline $D=\{a1,b1,c1,d1,b2,c2\}$, with \newline $B=\{b1,c1,d1,c2\}$, 
  \qquad $C=\{b1,c1\}$. \newline The order $C\prec D$ can easily be seen to follow from IIa and IIb.
 \item[(IVb)] 
 $A=\begin{picture}(31,30)(0,0)
  \curve(3,3, 7,7)
  \curve(3,7, 7,3)
  \curve(10,0, 20,0)
  \curve(0,10, 30,10)
  \curve(0,20, 30,20)
  \curve(10,30, 30,30)
  \curve(0,10, 0,20)
  \curve(10,0, 10,30)
  \curve(20,0, 20,30)
  \curve(30,10, 30,30)
 \end {picture}
\prec
 D=\begin{picture}(40,40)(0,0)
  \curve(3,3, 7,7)
  \curve(3,7, 7,3)
  \curve(10,0, 30,0)
  \curve(0,10, 40,10)
  \curve(0,20, 40,20)
  \curve(0,30, 40,30)
  \curve(10,40, 30,40)
  \curve(0,10, 0,30)
  \curve(10,0, 10,40)
  \curve(20,0, 20,40)
  \curve(30,0, 30,40)
  \curve(40,10, 40,30)
 \end {picture}\quad$
with \quad
 $B=\begin{picture}(40,40)(0,0)
  \curve(3,3, 7,7)
  \curve(3,7, 7,3)
  \curve(20,0, 30,0)
  \curve(10,10, 40,10)
  \curve(0,20, 40,20)
  \curve(0,30, 40,30)
  \curve(10,40, 30,40)
  \curve(0,20, 0,30)
  \curve(10,10, 10,40)
  \curve(20,0, 20,40)
  \curve(30,0, 30,40)
  \curve(40,10, 40,30)
 \end {picture}$
and 
 $C=\begin{picture}(30,30)(0,0)
  \curve(3,3, 7,7)
  \curve(3,7, 7,3)
  \curve(10,10, 30,10)
  \curve(10,20, 30,20)
  \curve(10,30, 30,30)
  \curve(10,10, 10,30)
  \curve(20,10, 20,30)
  \curve(30,10, 30,30)
 \end {picture}$ \newline
$A=\{a2,b1,b2,b3,c2,c3\} \prec D$ with \newline 
 $D=\{a2,a3,b1,b2,b3,b4,c1,c2,c3,c4,d2,d3\}$, \newline
 $B=\{a3,b2,b3,b4,c1,c2,c3,c4,d2,d3\}$, \newline
  $C=\{b2,b3,c2,c3\}$. 
 \newline  
Relation $A\prec B$ is a consequence of axiom IIIa, using $B=A'\cup A''$, where $A'=\{b2,c1...\}$ 
 and also $A''=\{a3,b2...\}$ are
 translates of $A$, and $A'\cap A''=\{b2,c3\}\prec A$ by applications of IIa and IIb. 
 The relation $C\prec D$ can be proven with IIa and IIIa.
 \item[(V)] 
 $A=\begin{picture}(40,40)(0,0)
  \curve(3,3, 7,7)
  \curve(3,7, 7,3)
  \curve(20,0, 30,0)
  \curve(20,10, 30,10)
  \curve(0,20, 30,20)
  \curve(0,30, 40,30)
  \curve(30,40, 40,40)
  \curve(0,20, 0,30)
  \curve(10,20, 10,30)
  \curve(20,0, 20,30)
  \curve(30,0, 30,40)
  \curve(40,30, 40,40)
 \end {picture}
\prec
 D=\begin{picture}(50,50)(0,0)
  \curve(3,3, 7,7)
  \curve(3,7, 7,3)
  \curve(20,0, 40,0)
  \curve(20,10, 40,10)
  \curve(0,20, 40,20)
  \curve(0,30, 50,30)
  \curve(0,40, 50,40)
  \curve(30,50, 40,50)
  \curve(0,20, 0,40)
  \curve(10,20, 10,40)
  \curve(20,0, 20,40)
  \curve(30,0, 30,50)
  \curve(40,0, 40,50)
  \curve(50,30, 50,40)
 \end {picture}
\quad$
with
 $B=\begin{picture}(40,50)(0,0)
  \curve(3,3, 7,7)
  \curve(3,7, 7,3)
  \curve(20,0, 30,0)
  \curve(20,10, 30,10)
  \curve(0,20, 30,20)
  \curve(0,30, 40,30)
  \curve(0,40, 40,40)
  \curve(30,50, 40,50)
  \curve(0,20, 0,40)
  \curve(10,20, 10,40)
  \curve(20,0, 20,40)
  \curve(30,0, 30,50)
  \curve(40,30, 40,50)
 \end {picture}
\quad$
and 
 $C=
\begin{picture}(50,40)(0,0)
  \curve(3,3, 7,7)
  \curve(3,7, 7,3)
  \curve(20,0, 40,0)
  \curve(20,10, 40,10)
  \curve(0,20, 40,20)
  \curve(0,30, 50,30)
  \curve(30,40, 50,40)
  \curve(0,20, 0,30)
  \curve(10,20, 10,30)
  \curve(20,0, 20,30)
  \curve(30,0, 30,40)
  \curve(40,0, 40,40)
  \curve(50,30, 50,40)
 \end {picture}$ \newline
$A=\{a3,b3,c1,c2,c3,d4\} \prec D$ with \newline
 $D=\{a3,a4,b3,b4,c1,c2,c3,c4,d1...d5,e4\}$, \newline
 $B=\{a3,a4,b3,b4,c1...c4,d4,d5\}$, \newline $C=\{a3,b3,c1,c2,c3,d1...d4,e4\}$. \newline
 Relation $A\prec B$ is a consequence of axiom IVa, using $B=A\cup A'$, where $A'=\{a4,b4...\}$ 
 is a translate of $A$, and $A\cap A'=\{c2,c3\}\prec B$ by repeated application of IIa and IIb. 
 The analogue procedure proves $A\prec C$.
 \item[(VI)] 
 $A=\begin{picture}(30,30)(0,0)
  \curve(3,3, 7,7)
  \curve(3,7, 7,3)
  \curve(0,0, 10,0)
  \curve(0,10, 30,10)
  \curve(0,20, 30,20)
  \curve(20,30, 30,30)
  \curve(0,0, 0,20)
  \curve(10,0, 10,20)
  \curve(20,10, 20,30)
  \curve(30,10, 30,30)
 \end {picture}\quad
\prec\quad
 B=\begin{picture}(40,30)(0,0)
  \curve(3,3, 7,7)
  \curve(3,7, 7,3)
  \curve(0,0, 20,0)
  \curve(0,10, 40,10)
  \curve(0,20, 40,20)
  \curve(20,30, 40,30)
  \curve(0,0, 0,20)
  \curve(10,0, 10,20)
  \curve(20,0, 20,30)
  \curve(30,10, 30,30)
  \curve(40,10, 40,30)
 \end {picture}\quad$
$\prec\quad
 D=\begin{picture}(60,50)(0,0)
  \curve(3,3, 7,7)
  \curve(3,7, 7,3)
  \curve(0,0, 20,0)
  \curve(0,10, 40,10)
  \curve(0,20, 40,20)
  \curve(20,30, 60,30)
  \curve(20,40, 60,40)
  \curve(40,50, 60,50)
  \curve(0,0, 0,20)
  \curve(10,0, 10,20)
  \curve(20,0, 20,40)
  \curve(30,10, 30,40)
  \curve(40,10, 40,50)
  \curve(50,30, 50,50)
  \curve(60,30, 60,50)
 \end {picture}$ \newline
$A=\{a1,a2,b2,c2,c3\} \prec D$, with \newline
 $D=\{a1,a2,b1,b2,c2,c3,c4,d2,d3,d4,e4,e5,f4,f5\}$,\newline and with the intermediate set 
 \newline $B=\{a1,a2,b1,b2,c2,c3,d2,d3\}$. \newline 
 The relation $A \prec B$ is established with IVa, because of $B=A\cup A'$, where $A'=\{b1,b2...\}$ 
 is a translate of $A$, and $A\cap A'\prec B$.
 Relation $B \prec D$ is established with either IIIa or IVa, because of $D=B\cup B'$, 
 where $B'=\{c3,c4...\}$ is a translate of $B$, and $B\cap B'\prec B$ or 
 $B\cap B'\prec D$. 
\end{enumerate}

To see the necessity of the axioms, their logical independence, 
one has to see in each case, that the order relation 
does not follow from the application of another axiom instead. 
Here Lemma \ref{unionlem} helps. It is easy in each example, to find out all the 
sets $A'$, $A''$, $\ldots$ which are eqivalent to $A$ and subsets of $D$, and then to test 
all the chains $A \prec ? A'\cup A'' \prec ? \ldots \prec ? D$.
For the examples to IIIb and IVb there are several ordered chains, but 
one has always to use the special axiom which is just under inspection.

\section{Monotonicity of mean entropy}\label{monmean}

\begin{thm}[Monotone decrease]\label{monotoneth}
Assume that equivalent sets have the same mean entropy. 
Then mean entropy is monotone decreasing:
\beq\label{monotone}
A\prec D \Rightarrow s(A) \geq s(D)
\eeq
\end{thm}
\begin{proof}
We write the strong subadditivity of the entropy as a relation for the mean entropy:
\beq
s(A\cup B) \leq \lambda_A s(A)+\lambda_B s(B) -\lambda_C s(C)
\eeq
where we denote $A\cap B=C$, as in the axioms II, III and IV, and use 
\beq
\lambda_A = \frac{\mu (A)}{\mu (A\cup B)}, \qquad 
\lambda_B = \frac{\mu (B)}{\mu (A\cup B)}, \qquad 
\lambda_C = \frac{\mu (C)}{\mu (A\cup B)}.
\eeq
Note that $0 \leq \lambda_C \leq \min \{\lambda_A,\lambda_B\} \leq 1$ 
and $\lambda_A +\lambda_B = 1+\lambda_C$. 
In case $A\cap B=\emptyset$, we define $\lambda_C s(C=\emptyset ) =0$, in accordance with the 
simple subadditivity of the entropy (which would be sufficient to check axiom II),
\beq
S(A\cup B) \leq S(A)+S(B).
\eeq
Now we proceed in an inductive way along the process of creating the ordering, similar 
as we did in the proof of the Theorem \ref{twobe}. We check each axiom, and assume 
that the inequality (\ref{monotone}) holds for the pairs of sets which are assumed to be in 
an order relation already. 
\begin{enumerate}
\item[ (I)]
 Equivalent sets have the same mean entropy, so the inequalities are invariant.
\item[(II)]
 We have $\lambda_A+\lambda_B=1$, $s(B)\leq s(A)$, hence
 $s(D) \leq \lambda_A s(A)+\lambda_B s(B) \leq s(A)$.
\item[(III)]
 Now $s(C)\geq s(B)$, $s(B)\leq s(A)$, so
 $s(D) \leq \lambda_A s(A)+\lambda_B s(B) -\lambda_C s(C) \leq \lambda_A s(A)+
(\lambda_B -\lambda_C )s(B) \leq (\lambda_A +\lambda_B-\lambda_C) s(A)=s(A)$.
\item[(IV)]
 Here $s(C)\geq s(D)$, implying 
 $s(D) \leq \lambda_A s(A)+\lambda_B s(B) -\lambda_C s(D)$ $\Rightarrow$
 $s(D) \leq (\lambda_A s(A)+\lambda_B s(B))/(1+\lambda_C)$,  and $s(B)\leq s(A)$
 $\Rightarrow$ $s(D)\leq s(A)$.
\item[(V)]
 Here we have $s(D=B\cup C) \leq \lambda_B s(B)+\lambda_C s(C) -\lambda_A s(A)$, 
 $s(B)\leq s(A)$, $s(C)\leq s(A)$ and $\lambda_B +\lambda_C = 1+\lambda_A$, which 
 obviously implies $s(D)\leq s(A)$.
\item[(VI)]
 Transitivity is obvious.
\end{enumerate}
\end{proof}
It occurs as a surprise, that we had not to assume equivalent sets to have the same measure,
as it is naturally the case in all our examples.

\section{Octogonal sets in a two-dimensional lattice}\label{octogons}

We extend the monotonicity of mean entropy, known for rectangular boxes, to the 
family $\oo$ of convex octogonal sets. $\oo$ is defined as consisting of all those 
finite convex subsets of $\Z^2$, 
whose boundaries are horizontal, vertical and diagonal lines, with some exceptions:  
Excluded are oblique rectangles. (But oblique straight lines are elements of $\oo$.) 
We characterize each set $A$ in $\oo$ by the coordinates of the most left point 
in the lowest line, and by a sequence of eight non negative integers, 
$(m,n,p,\ldots u)$, which describe its boundary $\bb_A$. The length of the lower 
horizontal boundary is $m$, the other boundary lines follow counter clockwise, with 
the lengths $n\sqrt{2}, p, \ldots u\sqrt{2}$. The following drawings show a set $A$, 
each point in $\Z^2$ represented by a square, and its boundary $\bb_A =(4,1,2,2,1,3,1,1)$: 
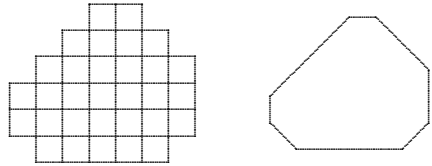
\begin{center}
$\begin{picture}(70,60)(0,0)
  \curve(10,0, 60,0)
  \curve(0,10, 70,10)
  \curve(0,20, 70,20)
  \curve(0,30, 70,30)
  \curve(10,40, 70,40)
  \curve(20,50, 60,50)
  \curve(30,60, 50,60)
  \curve(0,10, 0,30)
  \curve(10,0, 10,40)
  \curve(20,0, 20,50)
  \curve(30,0, 30,60)
  \curve(40,0, 40,60)
  \curve(50,0, 50,60)
  \curve(60,0, 60,50)
  \curve(70,10, 70,40)
 \end {picture} \qquad 
\begin{picture}(60,50)(0,0)
  \curve(15,5, 55,5)
  \curve(55,5, 65,15)
  \curve(65,15, 65,35)
  \curve(65,35, 45,55)
  \curve(45,55, 35,55)
  \curve(35,55, 5,25)
  \curve(5,25, 5,15)
  \curve(5,15, 15,5)
 \end {picture}$
\end{center} 
Note that the boundary lines are included in the set. 

Since the boundary is a closed curve, the parameters have to obey 
\beqa
u+m+n&=&q+r+s\label{closed1},\\
n+p+q&=&s+t+u\label{closed2},
\eeqa
and for {\it each} sequence of 8 non negative numbers obeing these relations there 
exists a boundary. 
The length of any boundary line (or of more boundary lines) may be zero. 
Then two (or more) corners, otherwise at the endpoints of this boundary line, 
coalesce into one. Octogons which are ``Lines'' have two pieces of boundary 
(representing two opposite directions of the boundary curve). 
A special boundary is $\bb_E=(0,0,\ldots,0)$, characterizing elementary sets, 
the ``atoms''.

Again we define a set $A'$ to be equivalent to $A$, if and only if $A'$ is a translate of $A$.
An equivalence class of octogons is obviously characterized by the boundary, 
which is common to all of the members of the equivalence class. 
Moreover, and to show this is the goal of this section, 
the ordering of the octogons corresponds to an ordering of the boundaries.
\begin{define}[Ordering of boundaries]\label{orderbd}
Boundary $\bb_A =(m_A, \ldots u_A)$ is said to be piecewise shorter than 
$\bb_B =(m_B, \ldots u_B)$, denoted as $\bb_A < \bb_B$, if $\bb_A \not= \bb_B$ and 
$m_A \leq m_B ,n_A\leq n_B, \ldots u_A \leq u_B$.
\end{define}
\begin{thm}[Ordering of octogons]\label{octogonorder}
Two octogons, $A$ and $D$, elements of $\oo$, are in the order relation $A\prec D$, 
if and only if $\bb_A < \bb_D$.
\end{thm}
This ordering includes {\it blow-up sequences}: Let $A_\nu$ be characterized by the 
boundary $\nu \bb_A =(\nu m,\nu n, \ldots \nu u)$, with $\nu$ a positive integer. 
Then $\nu <\pi$ implies $A_\nu<A_\mu$. 
\begin{proof}
One direction, assuming $A\prec D$, is easy. By Lemma \ref{unionlem}, 
$D$ is a union of some translates of $A$. 
So each boundary line of $D$ contains the boundary line of at least one of these 
translated sets and can thus not be shorter. 
The proof of the assertion in the other logical direction fills the remaining part 
of this section.
\end{proof}
Consider the circumference of a boundary
\beq\label{circumference}
\ell (D) = m_D +p_D+r_D+t_D+\sqrt{2}(n_D+q_D+s_D+u_D).
\eeq
Each finite interval in $\R$ contains only a finite number of such lengths, so they can all
be numbered and ordered as $\ell_1 < \ell_2< \ldots <\ell_N<\ldots$. The proof will be by 
{\it induction on the circumference} $\ell (D)$, this is, by induction on N. 
Since $\ell(E)=0$ holds for the atoms only, with no other 
boundaries piecewise shorter than $\bb_E$ existing, there is nothing to prove at the start of 
the induction. 
The induction hypothesis is, that $\bb_A < \bb_B$ and 
$\ell (B) < \ell (D)$ imply already $A\prec B$.
Now, if $\bb_A < \bb_D$, and if, concerning the ordering, there are other boundaries in between, 
$\bb_A<\ldots\bb_B<\bb_D$, then $\ell(B)<\ell(D)$, and it remains to prove $B\prec D$ only. 
In other words, we assume that there are no other boundaries between $\bb_A$ and $\bb_D$. 

Since the linear relations (\ref{closed1}) and (\ref{closed2}) hold for both $\bb_A$ and $\bb_D$, 
they hold also for the piecewise difference $\bb_D -\bb_A = (m_D-m_A,\ldots)$, 
and this difference can be interpreted as boundary $\bb_M$ of a ``{\bf molecule}'' $M$, 
a set which is in the ordering one level directly above the atoms, with no other set 
in between. It is not difficult to find out that there are exactly twelve equivalence 
classes of molecules in $\oo$, which can be grouped into four types. Representatives of these 
types are:
\begin{center}
$M_1=\{a1,a2\}=\begin{picture}(20,10)(0,0)
  \curve(0,0, 20,0)
  \curve(0,10, 20,10)
  \curve(0,0, 0,10)
  \curve(10,0, 10,10)
  \curve(20,0, 20,10)
\end {picture}$\hspace{1pt} ,\quad
$M_2=\{a2,b1\}=\begin{picture}(20,20)(0,0)
  \curve(10,0, 20,0)
  \curve(0,10, 20,10)
  \curve(0,20, 10,20)
  \curve(0,10, 0,20)
  \curve(10,0, 10,20)
  \curve(20,0, 20,10)
\end {picture}$\hspace{1pt},\end{center}
\begin{center}
$M_3=\{a1,a2,b1\}=\begin{picture}(20,20)(0,0)
  \curve(0,0, 20,0)
  \curve(0,10, 20,10)
  \curve(0,20, 10,20)
  \curve(0,0, 0,20)
  \curve(10,0, 10,20)
  \curve(20,0, 20,10)
\end {picture}$\hspace{1pt},\quad
$M_4=\{a1,b1,b2,c1\}=\begin{picture}(30,20)(0,0)
  \curve(0,0, 30,0)
  \curve(0,10, 30,10)
  \curve(10,20, 20,20)
  \curve(0,0, 0,10)
  \curve(10,0, 10,20)
  \curve(20,0, 20,20)
  \curve(30,0, 30,10)
\end {picture}$\hspace{1pt}. 
\end{center}
Each of these sets represents an equivalence class, and, together with the equivalence classes 
defined by its one rotated version or three rotated versions, it represents a type. 

There is a relation between the sets $M_i$ and the process of constructing $D$ out of $A$.
\begin{define}[Convolution of sets]
Let $A=\{\aaa_i, 1\leq i\leq I\}$, $C=\{\cc_j, 1\leq j\leq J\}$, then the convolution of 
$A$ with $C$ is 
\beq
A\ast C=\{\aaa_i+\cc_j, 1\leq i\leq I, 1\leq j\leq J\}.
\eeq
\end{define}
\begin{lem}[Convolution is Addition]\label{convadd}
If $A$ and $M$ are octogonal sets in $\oo$, but not orthogonal oblique lines, then 
$D=A\ast M \in\oo$ and $\bb_D=\bb_A+\bb_M$.
\end{lem}
\begin{proof}
It is not difficult to see, that $D$ is a convex set. 
(This would not be true, if $A$ and $M$ were oblique orthogonal lines, so we 
excluded this case; and we have no oblique rectangular boxes in $\oo$.) 
Consider a boundary line $\{\aaa_i, i\in Line_A\}$ of $A$ and the corresponding 
boundary line $\{\x_j, j\in Line_M\}$ of $M$, f.e. the lower 
horizontal boundary lines, with lenghts $m_A$ and $m_M$. 
(If its length happens to be zero, it consists of one point, f.e.  
the lowest point in the set.) 
Then $\{\aaa_i+\x_j, i\in Line_A, j\in Line_M\}$ is the corresponding boundary line of $A\ast M$, 
and its length is $m_A+m_M$.
\end{proof}
The convolution of $A$ with an atom produces a translate of $A$,
\beq
A\ast \{\x_i\}=A^{(i)} \sim A,
\eeq
so, for $M=\bigcup_i\{\x_i\}$,
\beq
A\ast M=\bigcup_i A^{(i)}.
\eeq

Now we are ready to apply the axioms and perform the step of induction.
\begin{proposition}
Let $\bb_A <\bb_D$, and $A\ast M_i =D$, for $A\in \oo$ and one of the twelve molecules $M_i$. 
Assume that $\bb_C<\bb_B$ and $\ell(B)<\ell(D)$ together imply $C\prec B$. 
Then $D\in\oo$ and $A\prec D$ holds.
\end{proposition}
\begin{proof}
If $A$ is an atom, then $A\ast M_i= M_i$ or $A\ast M_i\sim M_i$, and $A\prec M_i$ by axiom IIa. 

For larger sets we have to treat different cases differently; 
but since the system is invariant under 90-degree rotations, it is sufficient to prove 
this proposition with the molecules $M_1 \ldots M_4$. 
With $(m,n, \ldots u)$ we denote the boundary of $A$. Assume that the point $a1$ is the origin. 
\subsubsection*{Convolution with $M_1:$}
$A\ast M_1$ is the 
union of $A$ with $A'$, where $A'$ is $A$ translated by one step to the right. 
We have to investigate, whether $C=A\cap A'$ is empty or whether it can be compared with $A$ or 
with $D$. This depends on the form of $A$ in the lowest and in the highest region: 
If $A$ is not a vertical or oblique line, we say that the bottom of $A$ is 
\begin{enumerate}
\item[$\diamond$] {\it flat}, if $m\geq 1$; 
\item[$\diamond$] {\it sharp}, if $m=0$, and either $u=0$ and $n\geq 1$ or vice versa;
\item[$\diamond$] {\it rectangular}, if $m=0$, $u\geq 1$ and $n\geq 1$. 
\end{enumerate}
Analogously we denote the top, with $u,m,n$ replaced by $q,r,s$. 
The following pictures show the sharp top of some $A$ and the corresponding top of $D$ 
with $C$ inside marked with crosses, and the same for some $A$ with a flat top. 
\beq \label{fig1}
A_{sharp}=  \begin{picture}(31,30)(0,0)
  \curve(0,10, 30,10)
  \curve(0,20, 20,20)
  \curve(0,30, 10,30)
  \curve(0,0, 0,30)
  \curve(10,0, 10,30)
  \curve(20,0, 20,20)
 \end{picture},
\quad
D=  \begin{picture}(41,30)(0,0)
  \curve(0,10, 40,10)
  \curve(0,20, 30,20)
  \curve(0,30, 20,30)
  \curve(0,0, 0,30)
  \curve(10,0, 10,30)
  \curve(20,0, 20,30)
  \curve(30,0, 30,20)
  \curve(13,3, 17,7)
  \curve(13,7, 17,3)
  \curve(23,3, 27,7)
  \curve(23,7, 27,3)
  \curve(13,13, 17,17)
  \curve(13,17, 17,13)
 \end {picture};
\qquad
A_{flat}=  \begin{picture}(41,20)(0,0)
  \curve(0,10, 40,10)
  \curve(0,20, 30,20)
  \curve(0,0, 0,20)
  \curve(10,0, 10,20)
  \curve(20,0, 20,20)
  \curve(30,0, 30,20)
 \end{picture},
\quad
D=  \begin{picture}(51,20)(0,0)
  \curve(0,10, 50,10)
  \curve(0,20, 40,20)
  \curve(0,0, 0,20)
  \curve(10,0, 10,20)
  \curve(20,0, 20,20)
  \curve(30,0, 30,20)
  \curve(40,0, 40,20)
  \curve(13,3, 17,7)
  \curve(13,7, 17,3)
  \curve(23,3, 27,7)
  \curve(23,7, 27,3)
  \curve(33,3, 37,7)
  \curve(33,7, 37,3)
  \curve(13,13, 17,17)
  \curve(13,17, 17,13)
  \curve(23,13, 27,17)
  \curve(23,17, 27,13)
 \end {picture}.
\eeq
\begin{enumerate}
\item[$\bullet$]
{\it $A$ is a vertical or oblique line:} $A\cap A'=\emptyset$ and $A\prec D$ by axiom IIa. 
\item[$\bullet$]
{\it Top and bottom are sharp:} $A$ is either a parallelepiped, 
or an equilateral triangle, where the vertical line is the ``base'', or a trapezoid, 
also with the vertical line as the base. 
We have $m_C=0$ and also $q_C=0$, all the nonvanishing adjacent border lines of $C$ 
are shorter than 
the corresponding border lines of $A$, so $C$ can be compared with $A$: 
$\bb_C < \bb_A$, which implies $C\prec A$. 
With axiom IIIa we infer $A\prec D$. 
\newline
Example:  $A=\begin{picture}(21,45)(0,0)
  \curve(0,0, 10,0)
  \curve(0,10, 20,10)
  \curve(0,20, 20,20)
  \curve(0,30, 20,30)
  \curve(0,40, 10,40)
  \curve(0,0, 0,40)
  \curve(10,0, 10,40)
  \curve(20,10, 20,30)
 \end {picture}
\prec
 D=\begin{picture}(31,45)(0,0)
  \curve(13,13, 17,17)
  \curve(13,17, 17,13)
  \curve(13,23, 17,27)
  \curve(13,27, 17,23)
  \curve(0,0, 20,0)
  \curve(0,10, 30,10)
  \curve(0,20, 30,20)
  \curve(0,30, 30,30)
  \curve(0,40, 20,40)
  \curve(0,0, 0,40)
  \curve(10,0, 10,40)
  \curve(20,0, 20,40)
  \curve(30,10, 30,30)
 \end {picture}$, with $C$ inside.
\item[$\bullet$]
{\it Bottom is flat, top is sharp, or vice versa:} 
If $m\geq 1$, then $m_C = m-1$; the top of $C$ is sharp with adjacent border lines shorter 
than those at $A$, as in the above case. Other border lines of $C$ 
have the same lengths as the corresponding lines of $A$. So $C\prec A$, 
and we use again axiom IIIa. 
If $m=0$ but $r\geq 1$, the same argument is true, with top and bottom exchanged. 
Example:  $A=\begin{picture}(31,45)(0,0)
  \curve(0,0, 20,0)
  \curve(0,10, 30,10)
  \curve(0,20, 30,20)
  \curve(0,30, 20,30)
  \curve(0,40, 10,40)
  \curve(0,0, 0,40)
  \curve(10,0, 10,40)
  \curve(20,0, 20,30)
  \curve(30,10, 30,20)
 \end {picture}
\prec
 D=\begin{picture}(41,45)(0,0)
  \curve(13,3, 17,7)
  \curve(13,7, 17,3)
  \curve(13,13, 17,17)
  \curve(13,17, 17,13)
  \curve(13,23, 17,27)
  \curve(13,27, 17,23)
  \curve(23,13, 27,17)
  \curve(23,17, 27,13)
  \curve(0,0, 30,0)
  \curve(0,10, 40,10)
  \curve(0,20, 40,20)
  \curve(0,30, 30,30)
  \curve(0,40, 20,40)
  \curve(0,0, 0,40)
  \curve(10,0, 10,40)
  \curve(20,0, 20,40)
  \curve(30,0, 30,30)
  \curve(40,10, 40,20)
 \end {picture}$, with $C$ inside.
\item[$\bullet$]
{\it Top and bottom are flat:} 
(Horizontal lines and rectangles are here included.) 
Both $m_C=m-1$ and $r_C=r-1$ hold, with the other border lines of $C$ with the same 
lengths as those of $A$. So again $C\prec A$ holds and axiom IIIa is applicable. 
\newline 
Example:  $A=\begin{picture}(41,25)(0,0)
  \curve(0,0, 40,0)
  \curve(0,10, 40,10)
  \curve(10,20, 30,20)
  \curve(0,0, 0,10)
  \curve(10,0, 10,20)
  \curve(20,0, 20,20)
  \curve(30,0, 30,20)
  \curve(40,0, 40,10)
 \end {picture}
\prec
 D=\begin{picture}(51,25)(0,0)
  \curve(13,3, 17,7)
  \curve(13,7, 17,3)
  \curve(23,3, 27,7)
  \curve(23,7, 27,3)
  \curve(23,13, 27,17)
  \curve(23,17, 27,13)
  \curve(33,3, 37,7)
  \curve(33,7, 37,3)
  \curve(0,0, 50,0)
  \curve(0,10, 50,10)
  \curve(10,20, 40,20)
  \curve(0,0, 0,10)
  \curve(10,0, 10,20)
  \curve(20,0, 20,20)
  \curve(30,0, 30,20)
  \curve(40,0, 40,20)
  \curve(50,0, 50,10)
 \end {picture}\quad$, with $C$ inside. 
\newline
Now, if $m=r=1$ and $p=t=0$, the set $C$ is an oblique rectangle, not in $\oo$. 
It is nevertheless true, that $C\prec A$, as is shown under the item 
``Top and bottom are rectangular''. 
\end{enumerate}

If one of the extremal regions is rectangular, $C$ can not be compared with $A$. 
Either $m_C=1$ while $m=0$, or $r_C =1$ while $r=0$, or both. 
The picture shows the rectangular top of some $A$, 
the top of $D$ with $C$ inside, $C$ marked with crosses, 
and the top of a set $F$ which is needed in the proof. 
\beq\label{fig2}
A_{rectangular}=  \begin{picture}(51,30)(0,0)
  \curve(0,10, 50,10)
  \curve(10,20, 40,20)
  \curve(20,30, 30,30)
  \curve(0,0, 0,10)
  \curve(10,0, 10,20)
  \curve(20,0, 20,30)
  \curve(30,0, 30,30)
  \curve(40,0, 40,20)
 \end{picture},
\qquad
D=  \begin{picture}(61,30)(0,0)
  \curve(0,10, 60,10)
  \curve(10,20, 50,20)
  \curve(20,30, 40,30)
  \curve(0,0, 0,10)
  \curve(10,0, 10,20)
  \curve(20,0, 20,30)
  \curve(30,0, 30,30)
  \curve(40,0, 40,30)
  \curve(50,0, 50,20)
  \curve(13,3, 17,7)
  \curve(13,7, 17,3)
  \curve(23,3, 27,7)
  \curve(23,7, 27,3)
  \curve(33,3, 37,7)
  \curve(33,7, 37,3)
  \curve(43,3, 47,7)
  \curve(43,7, 47,3)
  \curve(23,13, 27,17)
  \curve(23,17, 27,13)
  \curve(33,13, 37,17)
  \curve(33,17, 37,13)
 \end {picture},
\qquad
F=  \begin{picture}(61,20)(0,0)
  \curve(0,10, 60,10)
  \curve(10,20, 50,20)
  \curve(0,0, 0,10)
  \curve(10,0, 10,20)
  \curve(20,0, 20,20)
  \curve(30,0, 30,20)
  \curve(40,0, 40,20)
  \curve(50,0, 50,20)
  \curve(13,3, 17,7)
  \curve(13,7, 17,3)
  \curve(23,3, 27,7)
  \curve(23,7, 27,3)
  \curve(33,3, 37,7)
  \curve(33,7, 37,3)
  \curve(43,3, 47,7)
  \curve(43,7, 47,3)
  \curve(23,13, 27,17)
  \curve(23,17, 27,13)
  \curve(33,13, 37,17)
  \curve(33,17, 37,13)
 \end {picture}.
\eeq
So, to prove $A\prec D$ one has to use axiom IVa, and establish the ordering $C\prec D$ first. 
In any case one has, to establish $C\prec D$, to construct a set $F\in \oo$, 
with $F\subset D$, $\ell (F)<\ell (D)$, and $\bb_C<\bb_F$, so that 
$C\prec F$ by assumption (the induction hypothesis to prove the theorem). 
Then one has to use a set $C'  \sim C$, such that $D=C'  \cup F$ and 
$G=C'  \cap F \prec F$ 
(or $G=\emptyset$), which implies $C \prec D$ by axiom IIIb (or by axiom IIb). 
\begin{enumerate}
\item[$\bullet$]
{\it Top is rectangular, bottom flat (or vice versa):} 
$F$ is the union $C^-\cup C\cup C^+$, 
where $C^-$ is $C$ translated one unit to the left, $C^+$ is $C$ translated one unit to the right. 
So $F= C\ast \{z1,a1,b1\}$, with $z1$ denoting the point left of $a1$. 
We discuss the case with the rectangular top, the other case is analogous by reflection. 
We move $C$ one unit upwards, denote this set as $C' $, 
and observe $D= F\cup C' $, see figure (\ref{fig2}).
The simplest case is just that of the example to axiom IVa in Section \ref{examples}. 
In this case $F\cap C' =\emptyset$, so axiom IIb makes the proof of $C \prec D$ complete. 
For any larger $A$ also $C$ is larger and  $G=F\cap C' \not=\emptyset$. 
By inspection of figure (\ref{fig2}) one sees that $r_G=r_F=3$, $q_G=q_C-1$, $s_G=s_C-1$. 
The other border lines have the same lengths as those of $C$. 
Since $\bb_C<\bb_F$, we have $\bb_G<\bb_F$, and $G\prec F$.
\newline
Example:
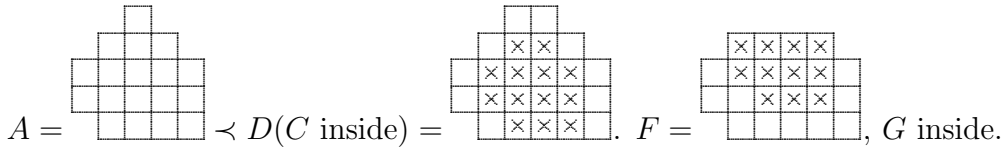
\newline
  $A=\begin{picture}(50,53)(0,0)
  \curve(10,0, 50,0)
  \curve(0,10, 50,10)
  \curve(0,20, 50,20)
  \curve(0,30, 50,30)
  \curve(10,40, 40,40)
  \curve(20,50, 30,50)
  \curve(0,10, 0,30)
  \curve(10,0, 10,40)
  \curve(20,0, 20,50)
  \curve(30,0, 30,50)
  \curve(40,0, 40,40)
  \curve(50,0, 50,30)
 \end {picture}
\prec
 D(C$ inside$)=\begin{picture}(61,53)(0,0)
  \curve(13,13, 17,17)
  \curve(13,17, 17,13)
  \curve(13,23, 17,27)
  \curve(13,27, 17,23)
  \curve(23,3, 27,7)
  \curve(23,7, 27,3)
  \curve(23,13, 27,17)
  \curve(23,17, 27,13)
  \curve(23,23, 27,27)
  \curve(23,27, 27,23)
  \curve(23,33, 27,37)
  \curve(23,37, 27,33)
  \curve(33,3, 37,7)
  \curve(33,7, 37,3)
  \curve(33,13, 37,17)
  \curve(33,17, 37,13)
  \curve(33,23, 37,27)
  \curve(33,27, 37,23)
  \curve(33,33, 37,37)
  \curve(33,37, 37,33)
  \curve(43,3, 47,7)
  \curve(43,7, 47,3)
  \curve(43,13, 47,17)
  \curve(43,17, 47,13)
  \curve(43,23, 47,27)
  \curve(43,27, 47,23)
  \curve(10,0, 60,0)
  \curve(0,10, 60,10)
  \curve(0,20, 60,20)
  \curve(0,30, 60,30)
  \curve(10,40, 50,40)
  \curve(20,50, 40,50)
  \curve(0,10, 0,30)
  \curve(10,0, 10,40)
  \curve(20,0, 20,50)
  \curve(30,0, 30,50)
  \curve(40,0, 40,50)
  \curve(50,0, 50,40)
  \curve(60,0, 60,30)
 \end {picture}$. 
$F=\begin{picture}(61,33)(0,0)
  \curve(13,23, 17,27)
  \curve(13,27, 17,23)
  \curve(13,33, 17,37)
  \curve(13,37, 17,33)
  \curve(23,13, 27,17)
  \curve(23,17, 27,13)
  \curve(23,23, 27,27)
  \curve(23,27, 27,23)
  \curve(23,33, 27,37)
  \curve(23,37, 27,33)
  \curve(33,13, 37,17)
  \curve(33,17, 37,13)
  \curve(33,23, 37,27)
  \curve(33,27, 37,23)
  \curve(33,33, 37,37)
  \curve(33,37, 37,33)
  \curve(43,13, 47,17)
  \curve(43,17, 47,13)
  \curve(43,23, 47,27)
  \curve(43,27, 47,23)
  \curve(43,33, 47,37)
  \curve(43,37, 47,33)
  \curve(10,0, 60,0)
  \curve(0,10, 60,10)
  \curve(0,20, 60,20)
  \curve(0,30, 60,30)
  \curve(10,40, 50,40)
  \curve(0,10, 0,30)
  \curve(10,0, 10,40)
  \curve(20,0, 20,40)
  \curve(30,0, 30,40)
  \curve(40,0, 40,40)
  \curve(50,0, 50,40)
  \curve(60,0, 60,30)
 \end {picture}$, $G$ inside. 
\item[$\bullet$]
{\it Top is rectangular, bottom sharp (or vice versa):} 
$F=C\ast H$, where $H=\{z0,z1,a0,a1,b1\}$ ($a0$ denotes the point below $a1$, etc.) 
is equivalent to the bottom of $D$: 
\beq
{\it bottom}(D)=  \begin{picture}(30,20)(0,0)
  \curve(0,0, 20,0)
  \curve(0,10, 30,10)
  \curve(0,0, 0,20)
  \curve(10,0, 10,20)
  \curve(20,0, 20,20)
  \curve(30,10, 30,20)
  \curve(13,13, 17,17)
  \curve(13,17, 17,13)
 \end {picture}
\eeq
$F\in \oo$ since $C\in \oo$ and $H\in\oo$. The remaining procedure is as above: 
We construct $C' =C\ast \{a2\}$, $G=C' \cap F$, so that $D=C' \cup F$, 
and observe $r_G=r_F=3$, $q_G=q_C-1$, $s_G=s_C-1$. 
The other border lines have the same lengths as those of $C$. 
Since $\bb_C<\bb_F$, we have $\bb_G<\bb_F$, and $G\prec F$.
This implies $C\prec D$.
\newline
Example:
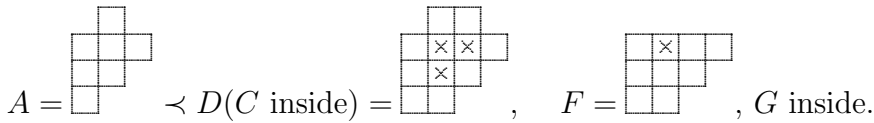
\newline
  $A=\begin{picture}(31,43)(0,0)
  \curve(0,0, 10,0)
  \curve(0,10, 20,10)
  \curve(0,20, 30,20)
  \curve(0,30, 30,30)
  \curve(10,40, 20,40)
  \curve(0,0, 0,30)
  \curve(10,0, 10,40)
  \curve(20,10, 20,40)
  \curve(30,20, 30,30)
 \end {picture}
\prec
 D(C$ inside$)=\begin{picture}(41,43)(0,0)
  \curve(13,13, 17,17)
  \curve(13,17, 17,13)
  \curve(13,23, 17,27)
  \curve(13,27, 17,23)
  \curve(23,23, 27,27)
  \curve(23,27, 27,23)
  \curve(0,0, 20,0)
  \curve(0,10, 30,10)
  \curve(0,20, 40,20)
  \curve(0,30, 40,30)
  \curve(10,40, 30,40)
  \curve(0,0, 0,30)
  \curve(10,0, 10,40)
  \curve(20,0, 20,40)
  \curve(30,10, 30,40)
  \curve(40,20, 40,30)
 \end {picture}$, \quad  
$F=\begin{picture}(41,33)(0,0)
  \curve(13,23, 17,27)
  \curve(13,27, 17,23)
  \curve(0,0, 20,0)
  \curve(0,10, 30,10)
  \curve(0,20, 40,20)
  \curve(0,30, 40,30)
  \curve(0,0, 0,30)
  \curve(10,0, 10,30)
  \curve(20,0, 20,30)
  \curve(30,10, 30,30)
  \curve(40,20, 40,30)
 \end {picture}$, $G$ inside. 
\item[$\bullet$]
{\it Top and bottom are rectangular:}
The bottom of $D$ with $C$ inside is like the bottom in figure (\ref{fig2}) upside down. 
$H$ is defined accordingly as $H=\{z1,a1,b1,a0\}\in \oo$.
It follows that $F=C\ast H\in\oo$, $C\prec F$. Again we use $C'=C\ast \{a2\}$, 
and perform the same procedures as in the last two cases. 
\newline
Example:
\newline
  $A=\begin{picture}(31,43)(0,0)
  \curve(10,0, 20,0)
  \curve(0,10, 30,10)
  \curve(0,20, 30,20)
  \curve(0,30, 30,30)
  \curve(10,40, 20,40)
  \curve(0,10, 0,30)
  \curve(10,0, 10,40)
  \curve(20,0, 20,40)
  \curve(30,10, 30,30)
 \end {picture}
\prec
 D(C$ inside$)=\begin{picture}(41,43)(0,0)
  \curve(13,13, 17,17)
  \curve(13,17, 17,13)
  \curve(13,23, 17,27)
  \curve(13,27, 17,23)
  \curve(23,13, 27,17)
  \curve(23,17, 27,13)
  \curve(23,23, 27,27)
  \curve(23,27, 27,23)
  \curve(10,0, 30,0)
  \curve(0,10, 40,10)
  \curve(0,20, 40,20)
  \curve(0,30, 40,30)
  \curve(10,40, 30,40)
  \curve(0,10, 0,30)
  \curve(10,0, 10,40)
  \curve(20,0, 20,40)
  \curve(30,0, 30,40)
  \curve(40,10, 40,30)
 \end {picture}$, \quad  
$F=\begin{picture}(41,33)(0,0)
  \curve(13,23, 17,27)
  \curve(13,27, 17,23)
  \curve(23,23, 27,27)
  \curve(23,27, 27,23)
  \curve(10,0, 30,0)
  \curve(0,10, 40,10)
  \curve(0,20, 40,20)
  \curve(0,30, 40,30)
  \curve(0,10, 0,30)
  \curve(10,0, 10,30)
  \curve(20,0, 20,30)
  \curve(30,0, 30,30)
  \curve(40,10, 40,30)
 \end {picture}$, $G$ inside. 
\end{enumerate}
Since it is needed at another place, we observe that the same procedure can be applied also 
if $A$ is an oblique rectangle, because also in this case $C\in\oo$ and $D\in\oo$. 
%
%
%
%
\subsubsection*{Convolution with $M_2:$}
$A\ast M_2$ is the union of $A'=A\ast \{b1\}$ with $A''=A\ast \{a2\}$. 

The case $A'\cap A''=\emptyset$ happens if $A$ is a horizontal or a vertical line, 
or a narrow oblique strip 
perpendicular to $M_2$, where the border of $A$ has $q=u=0$, and either $m=0$ and $t=1$ 
or $m=1$ and $t=0$. In these cases axiom IIa applies. 
Note that using oblique lines rectangular to $M_2$ is forbidden. 
(See the remark in the proof of Lemma \ref{convadd}.) 

If $C=A'\cap A''$ is not the empty set, it is always a precursor of $A$ in the ordering. 
To see this, one has to check the lower left end and the upper right end of $A$. 
We define the types for the lower left end, with obvious analogues for the 
upper right ends by symmetry. The pictures show again each type of end for some $A$, 
and the corresponding end of $D$ with $C$ inside marked with crosses. 
\begin{enumerate}
\item[$\bullet$]{\it sharp ends:}
Either $m\geq 2$, $s\geq 1$, $t=u=0$, so that $m_C=m-1$, $t_C=u_C=0$, 
\beq\label{fig3} 
A_{sharp}=  \begin{picture}(41,40)(0,0)
  \curve(0,0, 40,0)
  \curve(0,10, 40,10)
  \curve(10,20, 40,20)
  \curve(20,30, 40,30)
  \curve(0,0, 0,10)
  \curve(10,0, 10,20)
  \curve(20,0, 20,30)
  \curve(30,0, 30,40)
 \end{picture},
\qquad
D=  \begin{picture}(51,50)(0,0)
  \curve(10,0, 50,0)
  \curve(0,10, 50,10)
  \curve(0,20, 50,20)
  \curve(10,30, 50,30)
  \curve(20,40, 40,40)
  \curve(0,10, 0,20)
  \curve(10,0, 10,30)
  \curve(20,0, 20,30)
  \curve(20,0, 20,40)
  \curve(30,0, 30,50)
  \curve(40,0, 40,40)
  \curve(23,13, 27,17)
  \curve(23,17, 27,13)
  \curve(33,13, 37,17)
  \curve(33,17, 37,13)
  \curve(43,13, 47,17)
  \curve(43,17, 47,13)
  \curve(33,23, 37,27)
  \curve(33,27, 37,23)
  \curve(43,23, 47,27)
  \curve(43,27, 47,23)
  \curve(43,33, 47,37)
  \curve(43,37, 47,33)
 \end {picture},
\eeq
or $t\geq 2$, $n\geq 1$, $m=u=0$, and $t_C=t-1$, $m_C=u_C=0$. 

The length $s_C$ (or $n_C$) is also shorter than $s$ (or n); how much shorter it is depends 
also on the other end. 
\item[$\bullet$]
{\it rectangular ends:}
$u=0$, $m\geq 1$, $t\geq 1$, so $u_C=0$, $m_C=m-1$, $t_C=t-1$.
\beq
A_{rectangular}=  \begin{picture}(31,30)(0,0)
  \curve(0,0, 30,0)
  \curve(0,10, 30,10)
  \curve(0,20, 20,20)
  \curve(0,0, 0,30)
  \curve(10,0, 10,30)
  \curve(20,0, 20,20)
 \end{picture},
\quad
D=  \begin{picture}(41,40)(0,0)
  \curve(10,0, 40,0)
  \curve(0,10, 40,10)
  \curve(0,20, 30,20)
  \curve(0,30, 20,30)
  \curve(0,10, 0,40)
  \curve(10,0, 10,40)
  \curve(20,0, 20,30)
  \curve(30,0, 30,20)
  \curve(13,13, 17,17)
  \curve(13,17, 17,13)
  \curve(23,13, 27,17)
  \curve(23,17, 27,13)
  \curve(33,13, 37,17)
  \curve(33,17, 37,13)
  \curve(13,23, 17,27)
  \curve(13,27, 17,23)
  \curve(23,23, 27,27)
  \curve(23,27, 27,23)
  \curve(13,33, 17,37)
  \curve(13,37, 17,33)
 \end {picture}.
\eeq
\item[$\bullet$]
{\it flat ends:}
$u\geq 1$, so $u_C=u-1$, $m_C=m$, $t_C=t$.
\beq
A_{flat}=  \begin{picture}(31,30)(0,0)
  \curve(20,0, 30,0)
  \curve(10,10, 30,10)
  \curve(0,20, 20,20)
  \curve(0,20, 0,30)
  \curve(10,10, 10,30)
  \curve(20,0, 20,20)
 \end{picture},
\quad
D=  \begin{picture}(41,40)(0,0)
  \curve(30,0, 40,0)
  \curve(20,10, 40,10)
  \curve(10,20, 30,20)
  \curve(0,30, 20,30)
  \curve(0,30, 0,40)
  \curve(10,20, 10,40)
  \curve(20,10, 20,30)
  \curve(30,0, 30,20)
  \curve(23,13, 27,17)
  \curve(23,17, 27,13)
  \curve(33,13, 37,17)
  \curve(33,17, 37,13)
  \curve(13,23, 17,27)
  \curve(13,27, 17,23)
  \curve(23,23, 27,27)
  \curve(23,27, 27,23)
  \curve(13,33, 17,37)
  \curve(13,37, 17,33)
 \end {picture}.
\eeq
\end{enumerate}
If no end is sharp, both $n_C=n$ and $s_C=s$. Otherwise at least one of these border 
lines is shorter than those of $A$. In any case, $\bb_C< \bb_A$, so $C\prec A$, 
and axiom IIIa applies, proving $A\prec D$. 
\newline
Example:
\qquad 
 $A=
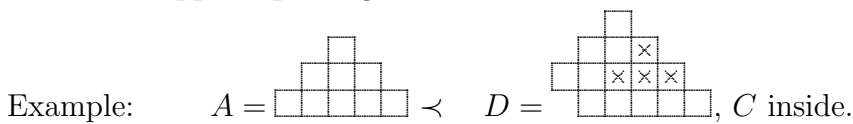
\begin{picture}(51,33)(0,0)
  \curve(0,0, 50,0)
  \curve(0,10, 50,10)
  \curve(10,20, 40,20)
  \curve(20,30, 30,30)
  \curve(0,0, 0,10)
  \curve(10,0, 10,20)
  \curve(20,0, 20,30)
  \curve(30,0, 30,30)
  \curve(40,0, 40,20)
  \curve(50,0, 50,10)
 \end {picture}
\prec \quad
 D=\begin{picture}(61,43)(0,0)
  \curve(23,13, 27,17)
  \curve(23,17, 27,13)
  \curve(33,13, 37,17)
  \curve(33,17, 37,13)
  \curve(33,23, 37,27)
  \curve(33,27, 37,23)
  \curve(43,13, 47,17)
  \curve(43,17, 47,13)
  \curve(10,0, 60,0)
  \curve(0,10, 60,10)
  \curve(0,20, 50,20)
  \curve(10,30, 40,30)
  \curve(20,40, 30,40)
  \curve(0,10, 0,20)
  \curve(10,0, 10,30)
  \curve(20,0, 20,40)
  \curve(30,0, 30,40)
  \curve(40,0, 40,30)
  \curve(50,0, 50,20)
  \curve(60,0, 60,10)
 \end {picture}$, $C$ inside. 
%
%
%
%
%
\subsubsection*{Convolution with $M_3:$}
\begin{enumerate}
\item[$\bullet$]
{\it Lines:} 
If $A$ is a vertical line, $(A\ast \{b1\})\cap (A\ast \{a1,a2\})=\emptyset$, 
and axiom IIb applies, using $D=(A\ast \{b1\})\cup (A\ast \{a1,a2\})$ and 
$A\prec (A\ast \{a1,a2\})$. 
For lines in other directions, the same argument applies, with the three points from $M_3$ 
appearing in appropriate permutations. 
\end{enumerate}
In the other cases, the types of the bottom and of the left side of $A$ dictate 
the way of proof,  
whether axiom IIIb can be used, or whether IVb is necessary. 
The types of the left side are defined in the same way as the types of the bottom, 
rotated by 90 degrees: {\it Flat} means $t\geq 1$, {\it rectangular} means 
$t=0$, $m\geq 1$, $u\geq 1$, and {\it sharp} means $t=0$ with either $s=0$, $r\geq 1$, 
$u\geq 1$, or $u=0$, $m\geq 1$, $s\geq 1$. 
\begin{enumerate}
\item[$\bullet$]
{\it Left side is flat:} $t\geq 1$. 
Define $B=A\ast M_1$. By assumption $A\prec B$, since $\ell (B)<\ell(D)$. 
Observe $D=A'\cup B$, with $A'=A\ast\{a2\}$. Then $C=A'\cap B \prec B$ has to be shown. 
If $s\geq 1$, then $s_C=s-1$, otherwise $t_C=t-1$. On the right side: 
If $q\geq 1$, then $q_C=q-1$; if $q=0$, $p\geq 1$, then $p_C=p-1$; 
if $p=q=0$, then $n\geq 1$, $n_C=n-1$. 
In all cases, $r_C\leq r$, so $\bb_C\leq \bb_B$ and $C\prec B$. 
By symmetry of the system and of the molecule $M_3$ under reflections at a 
diagonal, the proof for the cases with flat bottom, $m\geq 1$, is analogous. 
The example for the axiom IIIb is an example for this case. (What there has been termed $A$ 
is here $A'$:)
\item[$\bullet$]
{\it Left side is sharp:}
Since the cases with $m\geq 1$ can be treated in the way described above, we assume $m=0$. 
Also the lines have alresdy been considered, so we have here $u\geq 1$, $s=0$, $r\geq 1$. 
Also here $B=A\ast M_1$, $A'$ and $C$ defined as above. 
On the left side we observe $u_C=u-1$, on the right side the inequalities for different 
cases are as above. So again $C\prec B$ holds, and axiom IIIb applies. 
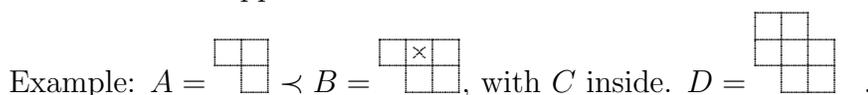
\newline 
Example:  $A=\begin{picture}(21,23)(0,0)
  \curve(10,0, 20,0)
  \curve(0,10, 20,10)
  \curve(0,20, 20,20)
  \curve(0,10, 0,20)
  \curve(10,0, 10,20)
  \curve(20,0, 20,20)
 \end {picture}
\prec
 B=\begin{picture}(31,23)(0,0)
  \curve(13,13, 17,17)
  \curve(13,17, 17,13)
  \curve(10,0, 30,0)
  \curve(0,10, 30,10)
  \curve(0,20, 30,20)
  \curve(0,10, 0,20)
  \curve(10,0, 10,20)
  \curve(20,0, 20,20)
  \curve(30,0, 30,20)
 \end {picture}$, with $C$ inside. 
 $D=\begin{picture}(41,33)(0,0)
  \curve(10,0, 30,0)
  \curve(0,10, 30,10)
  \curve(0,20, 30,20)
  \curve(0,30, 20,30)
  \curve(0,10, 0,30)
  \curve(10,0, 10,30)
  \curve(20,0, 20,30)
  \curve(30,0, 30,20)
 \end {picture}$.
\item[$\bullet$]
{\it Left side and bottom are rectangular:}
It is necessary to use $B=A\ast M_2$. Also in this case $A\prec B$. We have to investigate 
on $C=A\cap B$. The left side and the bottom are: $t_C=1$, $m_C=1$, $u_C=u-1$, 
$s_C=s-1$, $n_C=n-1$. The other border lines have the same lengths as in $A$. 
Form $F=C\ast H$, with $H=\{z1,a1,a2,b1\}=
\begin{picture}(31,23)(0,0)
  \curve(10,0, 20,10)
  \curve(10,10, 20,0)
  \curve(0,0, 30,0)
  \curve(0,10, 30,10)
  \curve(10,20, 20,20)
  \curve(0,0, 0,10)
  \curve(10,0, 10,20)
  \curve(20,0, 20,20)
  \curve(30,0, 30,10)
\end{picture}$. (The point $\{a1\}$ is marked with a cross.) 
Represent $D=(C\ast\{a2\})\cup F$, observe $C\prec F$. So, by axiom IIIb, $C\prec D$, 
and axiom IVb implies $A\prec D$. 
The example for axiom IVb is an example for this case. 
\end{enumerate}
%
%
%
\subsubsection*{Convolution with $M_4:$}
Define $B=A\ast \{z1,a1,b1\}$, $A'=A\ast \{a2\}$. 
So $D=A'\cup B$. If $A$ is a horizontal line, $A'\cap B=\emptyset$ and axiom IIb applies. 
For the other cases it is easy to see $C=A'\cap B\prec B$: Observe $r_C\leq r_B$. 
For the left side, consider the sequence $s,t,u$ of border lengths in $A$. The first one, 
which is not zero, is shorter by one in the set $C$. The analogue is true for the right side, 
considering $q,p,n$ and $q_C,p_C,n_C$. Obviously $m_C=m$. 
So $C\prec B$, and axiom IIIb applies. 
\end{proof}

\section{Remarks on continuum systems}
Consider a translation invariant state on the inductive limit of local algebras 
$\{\A_A\}$ on bounded convex measurable subsets $\{A\}$ of $\R^2$. 
Assume that the system fulfills the product property (\ref{product}) and the 
compatibility condition (\ref{compati}). Impose moreover a continuity condition, which 
enables an approximation of an octogon $A$ in $\R^2$ and the state restricted to 
$\{\A_A\}$ by sets $A_\nu$ and the states restricted to them, where the $A_\nu$ 
are unions of small elementary squares, as shown in the figures of 
Section \ref{octogons} in this paper. 
Then the statement of Theorem \ref{octogonorder} can be carried over to the octogons in $\R^2$. 
Also an extension to higher dimensions, establishing an ordering of 
some sets in $\Z^d$ and $\R^d$, seems to be possible. 

But these possible results are somehow unsatisfactory. 
Why just the octogons? 
If the monotonicity of mean entropy is true for sequences of octogons it is probably 
true for all kinds of convex sets. So I state now:
\begin{conjecture}[Blow up sequences]
Consider a system with all the properties stated above. 
Let $A$ be a convex bounded measurable set in $\R^d$, and define 
$\lambda A=\{\lambda \x; \quad \x\in A, \lambda>0\}$. 
Then $s(\lambda_1 A) \geq s(\lambda_2 A)$ if $\lambda_1 <\lambda_2$.
\end{conjecture}
The route to a proof of this statement might involve a stronger property: 
\begin{conjecture}[Ordering convex sets]
Consider a system with all the properties stated above. 
Let $A$ and $B$ be convex bounded measurable sets in $\R^d$, and define 
\beq
D=A \ast B=\{\x + {\bf y}; \quad \x \in A, {\bf y} \in B\},
\eeq
then mean entropy is decreasing, in the sense that $s(A)\geq s(D)$. 
\end{conjecture}

\section*{Appendix: Negative answers to the question of A. Kay and B. Kay}

The system is again a two-dimensional lattice. To avoid a discussion of the thermodynamic 
limit we consider some large box, with periodic boundary conditions for the interaction. 
The local Hilbert space for a single ``elementary'' site is $\C^2$. 
The system is now actually a classical discrete system, since we use only commuting 
operators: For each site $\alpha$ we use only the unit operator $\one$ and 
\beq
\sigma_\alpha \cong \left( \begin{array}{cc} 1 & 0 \\ 0 & -1 \end{array} \right).
\eeq
Then we use products:
\beq
\sigma_A=\left( \bigotimes_{\alpha \in A} \sigma_\alpha \right) \otimes
\left( \bigotimes_{\alpha \not\in A} {\one} \right) 
\eeq
Since the system is essentially a classical system, each state of such a system may 
be reinterpreted as a double periodic state 
of the infinite system on the lattice $\Z^2$: Any state is a probability measure 
on the set of configurations. Each configuration of the box can infinitely often be 
repeated periodically to give a configuration of the infinite system. Transfering a 
probability measure, which is invariant under translations of the box = torus, 
to this family of periodic configurations 
gives a translation invariant state of the infinite system.

We stay close to possible applications in physics, considering states defined 
by short range interactions, symmetric under translations, reflections and 90-degree rotations.
The states in the large box $G$ are represented by density matrices
\beq
\rho =e^{-\beta H}/\Tr e^{-\beta H},
\eeq 
$H$ a hamiltonian acting in $\Hh_G$, 
\beq\label{ham}
H=\sum_{A\in \G } \sigma_A,
\eeq
where $\G$ denotes any family of sets which is invariant under the symmetry group of 
periodic translations in the box $G$. The empty set is no element of $\G$. 

To enable a simple discussion of the restriction to a smaller set $D$, 
we consider very high temperatures, i.e. small $\beta$. 
The partition function $\Tr e^{-\beta H}$ and all the expectation values 
$\Tr \sigma_B \rho$ are analytic functions of $\beta$. 
\beq
Z(\beta)=\Tr e^{-\beta H} =Z(0)+O(\beta^2), \qquad with \quad Z(0)=2^{\mu (G)}
\eeq
\beq
\Tr \sigma_B \rho = -Z(0)^{-1}\beta\Tr\sigma_B H + O(\beta^2) \qquad for \quad B\not= \emptyset .
\eeq
Note that $\Tr\sigma_B\sigma_A=1$ if $B=A$, zero otherwise.  Therefore, 
from the Hamiltonian only the $\sigma_A$ with sets $A$ being subsets of $D$ contribute 
to the expectation values of $\sigma_B$ to first order in $\beta$, when $B\subset D$. 
So the restriction of the state to the region $D$ can be represented 
by a density matrix 
\beq
\rho_D=e^{-\beta H_D}/\Tr e^{-\beta H_D},
\eeq
with a hamiltonian $H_D$ acting in $\Hh_D$, 
\beq
H_D=\sum_{A\in\G (D)}\sigma_A + \tilde H_D,
\eeq
with $\| \tilde H_D \| =O(\beta )$, and where the family of
interaction-multiplets is denoted as 
\beq
\{A\in\G , A\subset D\} =\G (D).
\eeq
Now we can expand the mean entropy. The leading order is the same, $s_0=\ln 2$, for each set. 
The terms of order $\beta$ cancel, so we are interested in the terms of second order.
\beqa
s(D)&=& -\mu(D)^{-1}\Tr \rho_D\ln \rho_D \nonumber\\
&=& \mu(D)^{-1}\left( \ln\Tr e^{-\beta H_D} +\beta \Tr [H_D e^{-\beta H_D}]/
\Tr e^{-\beta H_D}\right) \nonumber\\
&=& \mu(D)^{-1}\left( \ln\Tr \one -\half \beta^2 \Tr H_D^2 /\Tr e^{-\beta H_D}\right) +O(\beta^3) \nonumber\\
&=& s_0 -\half \beta^2 \mu(D)^{-1}\Tr [\sum_{A\in\G (D)}\sigma_A]^2 /\Tr \one +O(\beta^3)
\eeqa
Observe, that $\Tr \sigma_A\sigma_{A'}=0$ for $A\not=A'$, and that $\sigma_A^2=\one$. 
So the calculation of the entropy $S(D)$ up to order $\beta^2$ amounts to counting 
how many interacting pairs or multiplets of sites are contained in the set $D$, 
that is, the number of sets in $\G (D)$, which we denote as 
\beq
n(D)= | \G (D)|.
\eeq
Mean entropy is thus 
\beq
s(D)=s_0-\half \beta^2n(D)/\mu (D) +O(\beta^3),
\eeq
and for small $\beta\not= 0$ 
\beq
s(B)<s(D) \quad \Longleftrightarrow \quad n(B)/\mu (B) > n(D)/\mu (D).
\eeq

Now we are ready to answer some questions. Consider the sets 
\newline $B=\{a1,a2,b1\}=
 \begin{picture}(20,23)(0,0)
  \curve(0,0, 20,0)
  \curve(0,10, 20,10)
  \curve(0,20, 10,20)
  \curve(0,0, 0,20)
  \curve(10,0, 10,20)
  \curve(20,0, 20,10)
\end {picture}$\hspace{1pt},
\qquad $C=\{a1,b1,c1\}=
 \begin{picture}(30,13)(0,0)
  \curve(0,0, 30,0)
  \curve(0,10, 30,10)
  \curve(0,0, 0,10)
  \curve(10,0, 10,10)
  \curve(20,0, 20,10)
  \curve(30,0, 30,10)
\end {picture}$\hspace{1pt},
\newline $D=\{a1,a2,b1,c1\}=
 \begin{picture}(30,23)(0,0)
  \curve(0,0, 30,0)
  \curve(0,10, 30,10)
  \curve(0,20, 10,20)
  \curve(0,0, 0,20)
  \curve(10,0, 10,20)
  \curve(20,0, 20,10)
  \curve(30,0, 30,10)
\end {picture}$\hspace{1pt}.
\newline
In comparing $B$ with $D$, we consider an interaction of diagonally nearest neighbors, 
$\G$ consisting of $\{a1,b2\}$,$\{a2,b1\}$ and its translates. 
Since $n(B)=n(D)=1$, we have $s(B)<s(D)$.

In comparing $C$ with $D$, consider $\G$ as consisting of triples like $C$ itself and 
its translates, and possibly also of the rotated triples and its translates. 
Since $n(C)=n(D)=1$, we have $s(C)<s(D)$.

The last example can easily be generalized to a lot of cases, where a larger set does 
{\it not} have lower mean entropy: If $C\subset D$, but no other set $C'$ equivalent 
to $C$ is a subset of $D$, consider $\G =\{C'\sim C\}$ as the set of interacting multiplets 
and the interaction Hamiltonian defined as in (\ref{ham}).

We may now extend the set of questions posed in \cite{ABK01} and ask: 
\newline {\bf Question:} 
{\it Considering two sets $C\subset D$, is it} {\rm necessary}, 
{\it that $D$ can be represented as a 
union of sets which are equivalent to $C$, to guarantee $s(C) \geq s(D)$ 
for any invariant state?}

This property, $D$ being a union of sets equivalent to $C$, 
is, alas, not sufficient. Here comes the counterexample promised after the 
proof of Lemma \ref{unionlem}:
Consider the sets 
\newline $C=\{a1,a2,b3,c1,c2\}=
 \begin{picture}(30,33)(0,0)
  \curve(0,0, 10,0)
  \curve(20,0, 30,0)
  \curve(0,10, 10,10)
  \curve(20,10, 30,10)
  \curve(0,20, 30,20)
  \curve(10,30, 20,30)
  \curve(0,0, 0,20)
  \curve(10,0, 10,30)
  \curve(20,0, 20,30)
  \curve(30,0, 30,20)
\end {picture}$\hspace{1pt}, 
\qquad and \quad 
$D=C\cup C'=
 \begin{picture}(50,33)(0,0)
  \curve(0,0, 10,0)
  \curve(0,10, 10,10)
  \curve(0,20, 50,20)
  \curve(20,0, 30,0)
  \curve(20,10, 30,10)
  \curve(10,30, 20,30)
  \curve(40,0, 50,0)
  \curve(40,10, 50,10)
  \curve(30,30, 40,30)
  \curve(0,0, 0,20)
  \curve(10,0, 10,30)
  \curve(20,0, 20,30)
  \curve(30,0, 30,30)
  \curve(40,0, 40,30)
  \curve(50,0, 50,20)
\end {picture}$\hspace{1pt},
\newline 
with $C'=\{c1,c2,d3,e1,e2\}$  being a translate of $C$. 
$\G$ consists of pairs of nearest neighbors, $A=\{a1,a2\}$ and its translates, 
possibly also of the rotated pairs, $B=\{a1,b1\}$ and its translates.
The comparison gives 
\beq \label{nmu}
\frac{n(C)}{\mu(C)}=\frac{2}{5} >\frac{n(D)}{\mu(D)}=\frac{3}{8}.
\eeq
So $s(C)<s(D)$ for small $\beta\not= 0$. As it has to be, 
because of Theorem \ref{monotoneth}, there is also no order 
relation between $C$ and $D$. This comes from the intersection $C\cap C'$ being in 
no order relation, neither with $C$, nor with $D$.
Moreover, this example can be extended. Translating $C$ again and again, 
always the same distances, one can form 
chains of different length, $D_N=C\cup C'\cup\ldots \cup C^{(N)}$. 
With the same method, calculating $n(D^{(N)})/\mu (D^{(N)})$, one can 
prove that $s(D^{(N)})$ is {\it increasing} in $N$.

With a state, which is {\it not} rotation invariant, it can be demonstrated, 
that oblique rectangles can not be ordered in the same way as the other orthogonal sets. 
Consider 
\newline $C=\{a2,b1,b2,b3,c2\}=
 \begin{picture}(31,33)(0,0)
  \curve(10,0, 20,0)
  \curve(0,10, 30,10)
  \curve(0,20, 30,20)
  \curve(10,30, 20,30)
  \curve(0,10, 0,20)
  \curve(10,0, 10,30)
  \curve(20,0, 20,30)
  \curve(30,10, 30,20)
\end {picture}$, \qquad  and \quad
 $D=C\cup C'=
 \begin{picture}(41,40)(0,0)
  \curve(20,0, 30,0)
  \curve(10,10, 40,10)
  \curve(0,20, 40,20)
  \curve(0,30, 30,30)
  \curve(10,40, 20,40)
  \curve(0,20, 0,30)
  \curve(10,10, 10,40)
  \curve(20,0, 20,40)
  \curve(30,0, 30,30)
  \curve(40,10, 40,20)
\end {picture}$,
\newline
with $C'=\{z2,a1,a2,a3,b2\}$, a translate of $C$. 
\newline 
$\G$ consists of pairs of diagonal neighbors, $A=\{a1,b2\}$ and its translates. 
The comparison gives again (\ref{nmu}) and $s(C)<s(D)$ for small $\beta\not= 0$.

\section*{Acknowledgements}
I thank Bernard Kay for for stimulating e-mail discussions, 
Hajime Moriya for telling me about the work in progress of H. Araki and himself, 
and I thank Elliott H. Lieb, Heide Narnhofer and Helmuth Urbantke for interest in this study.

\end{document}